\documentstyle[12pt,A4wide]{article}
\begin{document}

\rightline{Preprint DFPD 94/TH/67, December 1994.}

\vspace{0.5cm}

\begin{center}
{\bf STOCHASTIC VARIATIONAL APPROACH \\
     TO MINIMUM UNCERTAINTY STATES}
\end{center}

\begin{center}
Short title: STOCHASTIC VARIATIONAL \\
APPROACH TO MINIMUM UNCERTAINTY
\end{center}

\vspace{0.5cm}

\begin{center}
{\large Fabrizio Illuminati \footnote {E-Mail: illuminati@ipdgr4.pd.infn.it}
and Lorenza Viola \footnote {E-Mail: viola@mvxpd5.pd.infn.it}} \\

\vspace{0.3cm}

{\it Dipartimento di Fisica ``Galileo Galilei", \\
     Universit\`a di Padova, and INFN, Sezione di Padova, \\
     Via F. Marzolo 8, 35131 Padova, Italia}
\end{center}

\vspace{0.7cm}

\begin{abstract}
We introduce a new variational characterization of Gaussian diffusion
processes as minimum uncertainty states. We then define
a variational method
constrained by kinematics of diffusions and Schr\"{o}dinger
dynamics to seek states of local minimum uncertainty
for general non-harmonic potentials.
\end{abstract}

\vspace{0.5cm}


\vspace{0.5cm}

{\bf 1. Introduction}

\vspace{0.2cm}

Nelson stochastic quantization, originally proposed in 1966 \cite{nelson},
is currently recognized as an independent and self-consistent
formulation of quantum mechanics in the
language of stochastic processes.

The original Nelson scheme,
containing some {\it ad hoc} assumptions, has been later reconsidered
and conceptually clarified in the classic work by Guerra and Morato
\cite{guerramorato}. They proved that quantum dynamics can be obtained by
a stochastic variational principle, i.e. by
suitably extremizing the classical action along diffusive trajectories
replacing the deterministic ones. In this way stochastic variational
methods have creeped in into quantum physics. Both the practical
and the conceptual advantages of the variational scheme have since been
exploited in a number of different contexts \cite{guerramarra},
\cite{loffredomorato}, \cite{morato}.

In this work we introduce a variational method that yields
a new characterization of Gaussian diffusion processes. The
analysis is carried out for the one-dimensional case, but the results
can be easily extended in any number of dimensions. We prove
that a necessary and sufficient condition for a
diffusion process $q(t)$ to be Gaussian is that a certain
positive-defined functional of the probability
density $\rho(x,t)$ acquires a non-zero global minimum at each fixed
instant of time.
This functional is built by considering the variances
${\Delta q}^2$ of $q(t)$ and ${\Delta u}^2$ of the logarithmic derivative
$u(x,t) = \nu \partial_{x}[\ln\rho(x,t)]$, where $\nu$ is
the diffusion coefficient.
We then consider the ``uncertainty" product ${\Delta q}^2{\Delta u}^2$
as the quantity to be extremized against smooth variations
$\delta \rho$ of the probability density.

For the Nelson diffusions of stochastic mechanics
the quantity $u(x,t)$ is the so-called osmotic velocity, and
$\Delta q^{2}\Delta u^{2}$ is the osmotic uncertainty product.
One can then rephrase our result by saying that Nelson Gaussian diffusions
are all and only those that minimize the osmotic
uncertainty product at each fixed instant of time.

Historically, an uncertainty principle involving the diffusion coefficient had
been already conjectured by F\"{u}rth
in the study of Brownian Motion \cite{furth}, and
explicitly derived for the Nelson diffusions of stochastic mechanics
by exploiting Schwartz's inequality \cite{delapena}, \cite{defalco}.
Saturation of the osmotic uncertainty then yields the standard
harmonic-oscillator coherent and squeezed states \cite{demartino}.
Obviously, the same states can also be recovered in the stochastic
variational approach.

In fact, they are identified in two steps. Knowledge of the Gaussian
density allows first to derive their phase. This
is achieved by imposing Fokker-Planck kinematics through the continuity
equation. The latter connects the density with the gradient of the phase,
i.e. the so-called current velocity $v$. The harmonic-oscillator potential
is then determined by means of Schr\"{o}dinger dynamics in the
Hamilton-Jacobi-Madelung hydrodynamic form. Thus, in this scheme,
the kinematical and dynamical inputs are exploited {\it a posteriori},
once the osmotic uncertainty functional has been minimized.

On the other hand, in stochastic mechanics there is a current uncertainty
term $\Delta q^{2}\Delta v^{2}$ that has to be added to the osmotic
one to yield the full Heisenberg uncertainty product
$\Delta \hat{q}^{2}\Delta \hat{p}^{2}$. Minimization of the osmotic
term alone takes care of the current uncertainty
by fixing it to be either zero (Heisenberg minimum) or related
to the time-variation of the wave packet
spreading $\Delta q$ (Schr\"{o}dinger minimum).

It is however possible to consider a variational extremization
for the sum of both parts, which is in fact
a functional of the density and the current velocity. The
latter are not independent but are related by Fokker-Planck
kinematics and Schr\"{o}dinger dynamics, so that minimization
can be correctly performed only by including them
{\it a priori} as constraints in the variational procedure.
Furthermore, the osmotic uncertainty product $\Delta q^{2}\Delta u^{2}$
is a quantity defined at a fixed instant of time.
Upon integration over an arbitrary
time-interval $[t_{1},t_{2}]$, it is mapped into a dynamical functional
of the process (intuitively, a sort of uncertainty ``action").
The correct procedure to minimize this
dynamical osmotic uncertainty is then again to impose the
kinematical and dynamical constraints.

In the present work we first develop the free variational minimization
of the osmotic uncertainty, a result that holds for any diffusion process.
We then carry out the study for the various constrained variational schemes.

It is clear that the constrained approach to the minimum osmotic
uncertainty, by restricting
the class of allowed variations, should in principle
yield a larger class of solutions. In particular, it
should allow to assess the existence and to determine the relative minima
of the Heisenberg uncertainty product
as functions of the assigned external potentials.
Unfortunately, due to the intrinsic
mathematical complications,
we are at present able to carry out the analysis and present a complete
solution only for the kinematically constrained problem.

\vspace{0.5cm}

{\bf 2. Variational characterization of Gaussian diffusion processes}

\vspace{0.2cm}

Consider a one-dimensional diffusion process $q(t)$ with the associated
normalized probability density $\rho(x,t)$ and a diffusion
coefficient $\nu$ which may be constant or time-dependent.
Introduce the quantity
\begin{equation}
u(x,t) \; \equiv \; \nu\partial_{x}[\ln\rho(x,t)] \; = \;
\nu\frac{\partial_{x}\rho(x,t)}{\rho(x,t)} \, ,
\end{equation}

\noindent and consider the following functional of $\rho(x,t)$:
\begin{equation}
{{F}}[\rho] \; \equiv \; {\Delta q}^2 \,{\Delta u}^2 \, ,
\end{equation}

\noindent where
${\Delta q}^2\equiv Var\{q\}= \langle q^{2} \rangle - \langle q \rangle ^{2}$,
and
${\Delta u}^2 \equiv Var\{u\} = \langle u^{2} \rangle -\langle u \rangle ^{2}$
are the variances of $q(t)$ and $u(x,t)$ as functions of time. Here and in the
following $\langle \, \cdot \, \rangle$ denotes expectation with respect to the
density $\rho(x,t)$: for $u(x,t)$ as function of the process,
\begin{equation}
\langle u \rangle \; = \; \nu\int_{-\infty}^{+\infty}\partial_{x}\rho(x,t)dx
\; = \; 0 \, ,
\end{equation}

\noindent since the density vanishes at infinity. Thus $u(x,t)$ always
has, by construction, zero expectation value, and ${\Delta u}^2$
is simply $\langle u^{2} \rangle$.

We observe that ${F}$ is a smooth, positive-defined functional
of the density $\rho$. Pictorially, ${\Delta q}^2$ and ${\Delta u}^2$
are in competition: the first quantity measures the width of the
probability density, while the second measures its sharpeness;
we therefore expect on intuitive grounds that there exists an
absolute minimum of ${F}$. To verify this assertion
and to determine the explicit minimum value of the functional we extremize
${F}$ under smooth variations $\delta\rho$ of the probability density;
they are taken to be vanishing at the boundaries of integration.

Since minimization of ${F}$ must preserve normalization of
probability, the variations will be constrained by the requirement
that at each fixed instant of time $\int\rho(x,t)dx=1$. We thus consider
as the functional to be varied
\begin{equation}
{\cal {F}}[\rho] \; \equiv \; {F}[\rho]
+ \lambda(t)\int_{-\infty}^{+\infty}\rho(x,t)dx \, .
\end{equation}

\noindent The Lagrange multiplier $\lambda(t)$ shall be determined {\it
a posteriori} by taking the expectation of the variational equation.
The latter is obtained by imposing the vanishing of the variation
$\delta {\cal F}$ of the uncertainty functional:
\begin{equation}
\delta {\cal F} [\rho] \; \equiv \; \delta {F} [\rho] +
\lambda(t)\int_{-\infty}^{+\infty}\delta\rho(x,t)dx \; = \; 0 \, .
\end{equation}

Long but straightforward calculations lead to the following variational
equation for $u(x,t)$:
\begin{equation}
{\Delta u}^2\, (x^{2} -2\langle q \rangle x) - \left[ u^{2}(x,t) +
2\nu\partial_{x}u(x,t)\right] {\Delta q}^2 + \lambda(t) = 0 \, .
\end{equation}

\noindent Taking the expectation of both sides one finds
\begin{equation}
\lambda(t) = {\Delta u}^2 [\langle q \rangle ^{2} -2{\Delta q}^2] \, .
\end{equation}

\noindent Replacing the above expression for $\lambda$ in eq.(6)
gives the following inhomogeneous nonlinear differential
equation of the Riccati type for $u(x,t)$:
\begin{equation}
{\Delta q}^2 \left[ u^{2}(x,t) + 2\nu\partial_{x}u(x,t)\right] =
{\Delta u}^2 \, x^{2} -2{\Delta u}^2\, \langle q \rangle x  + {\Delta u}^2
[{\langle q \rangle }^{2} -2{\Delta q}^2] \, .
\end{equation}

The right hand-side is of the form $\alpha(t) x^{2}
+ \beta(t) x + \gamma(t)$: it follows that the solution $u(x,t)$, if
it exists, must be of the form $a(t)x + b(t)$.
The condition that $u(x,t)$ must have a vanishing expectation fixes
$b(t) = -a(t)\langle q \rangle$, so that the solution is finally of
the form
\begin{equation}
u(x,t) \; = \; a(t)[x - \langle q \rangle ] \, .
\end{equation}

The coefficient $a(t)$ is determined by the requirement that
$u$ satisfies identically the variational equation;
inserting expression (9) in eq.(8) and equating, respectively, the
coefficients of the powers $x^{2},x$ and $x^{0}$, yields $a(t) =
-\Delta u/\Delta q$ and
\begin{equation}
{\Delta q}^2\, {\Delta u}^2 \; = \; \nu^{2} \, .
\end{equation}

\noindent From the explicit form
\begin{equation}
u(x,t) \; = \; -\frac{\nu}{{\Delta q}^2}\left( x-\langle q\rangle \right) \, ,
\end{equation}

\noindent we readily trace back the expression for the normalized
probability density
\begin{equation}
\rho(x,t) \; = \; \frac{1}{ (2\pi)^{1/2} {\Delta q}}
\mbox{exp} \left\{ {-\frac{(x - \langle q \rangle )^{2} }{2 {\Delta q}^2}}
\right \} \, .
\end{equation}

It is possible to check, by convexity of the functional or by taking
second variations, that the extremum (10) is indeed a minimum.

In conclusion, we have proven the following

\noindent {\sl Theorem}:
A necessary and sufficient condition for a smooth
probability density $\rho(x,t)$ to be Gaussian is that it makes
stationary the functional $Var\{q\}\cdot Var\{u\}$, where $u \, = \,
\nu\partial_{x}[\ln\rho(x,t)]$ and $\nu$ is an arbitrary function
of time (for diffusion processes, the diffusion coefficient).
The extremum corresponds to a minimum with value $\nu^{2}$.

Sufficiency was just shown; necessity trivially follows from
the Gaussian characterization, eqs.(11)-(12), and from the definition
of ${{F}}[\rho]$.

\newpage

{\bf 3. Minimum uncertainty Nelson diffusions: coherent states}

\vspace{0.2cm}

The theorem proved in the previous section finds
application in the framework of stochastic formulation of quantum mechanics.
We recall that Nelson stochastic quantization \cite{guerranelson}
associates to each quantum state of a point particle of mass $m$
a configurational diffusion process $q(t)$ governed by Ito's stochastic
differential equation
\begin{equation}
dq(t) \; = \; v_{(+)}(q(t),t)dt + \left( \frac{\hbar}{2m}
\right)^{\frac{1}{2}} dw(t) \, , \; \; \; \; dt > 0 \, \, .
\end{equation}

\noindent Here $v_{(+)}(q(t),t)$ is the forward drift, $\hbar$ is
Planck constant, $\nu=\hbar/2m$
is the diffusion coefficient, and
$dw(t)$ is the time-increment of a Wiener process $w(t)$,
with expectation $\left\langle dw(t) \right\rangle =0$ and
covariance $\left\langle dw^{2}(t) \right\rangle =2\,dt$.
The forward and backward drifts, $v_{(+)}(x,t)$ and
$v_{(-)}(x,t)$, are defined by the conditional expectations
\[
v_{(+)}(x,t) \; \equiv \; \lim_{\Delta t \rightarrow 0^{+}}
\Big\langle \frac{q(t + \Delta t) - q(t)}{\Delta t}
\Big| q(t) = x \Big\rangle \, ,
\]
\begin{equation}
\end{equation}
\[
v_{(-)}(x,t) \; \equiv \; \lim_{\Delta t \rightarrow 0^{+}}
\Big\langle \frac{q(t) - q(t - \Delta t)}{\Delta t}
\Big|q(t) = x \Big\rangle \, .
\]

\noindent They represent respectively the mean forward (backward)
velocity fields.

In the hydrodynamic picture of the process, the drifts are replaced
by the osmotic velocity $u(x,t)$,
\begin{equation}
u(x,t) \; \equiv \; \frac{v_{(+)}(x,t) - v_{(-)}(x,t)}{2} \; = \;
\frac{\hbar}{2m}\partial_{x}[\ln\rho(x,t)] \, .
\end{equation}

\noindent and by the current velocity $v(x,t)$,
\begin{equation}
v(x,t) \; \equiv \; \frac{v_{(+)}(x,t) + v_{(-)}(x,t)}{2} \, .
\end{equation}

Finally, Fokker-Planck equation for the
probability density $\rho(x,t)$ takes the form of the
continuity equation
\begin{equation}
\partial_{t}\rho(x,t) \; = \; -\partial_{x}[\rho(x,t) v(x,t)] \, .
\end{equation}

At the dynamical level to each single-particle quantum state $\Psi(x,t)$
written in the hydrodynamic form
\begin{equation}
\Psi(x,t) \; = \; \sqrt{\rho(x,t)}\exp \left[ \frac{i}{\hbar}S(x,t) \right] \;
,
\end{equation}

\noindent
there corresponds in stochastic mechanics the diffusion process $q(t)$ with
\begin{equation}
\rho(x,t) \; = \; |\Psi(x,t)|^{2} \; ,
\end{equation}

\noindent and
\begin{equation}
v(x,t) \; = \; \frac{1}{m}\partial_{x}S(x,t) \; .
\end{equation}

The complex Schr\"odinger equation with potential $V(x,t)$ for the
wave function $\Psi$ is then equivalent to two coupled real
equations for the probability density $\rho$
and for the phase $S$ (or, alternatively, for the osmotic and current
velocities $u$ and $v$).
They are, respectively, the continuity equation (17)
and the gradient of the Hamilton-Jacobi-Madelung (HJM) equation which,
in terms of $u$ and $v$, reads
\begin{equation}
m\partial_{t}v + mv\partial_{x}v - mu\partial_{x}u -\frac{\hbar}{2}
\partial_{x}^{2}u \; = \; - \partial_{x}V(x,t) \; .
\end{equation}

The correspondence between expectations and uncertainties defined
in the stochastic and in the canonic formulations of quantum
mechanics are \cite{defalco}
\[
\langle \hat{q} \rangle_{\Psi} \; = \;
\langle q \rangle \, , \; \; \; \; \langle \hat{p}
\rangle_{\Psi} \; = \; m\langle v \rangle \, ,
\]

\begin{equation}
\Delta \hat{q} \; = \; \Delta q \, ,
\end{equation}

\[
\Delta \hat{p}^{2} \; = \; m^{2}[\Delta u^{2} + \Delta v^{2}] \, ,
\]

\noindent where $\hat{q}$ and $\hat{p}$ are the position and
momentum operators in the Schr\"odinger picture, $\langle \, \cdot \,
\rangle_{\Psi}$ denote the expectations of the operators in the
given state $\Psi$, $\langle \, \cdot \, \rangle$ denote the expectations
of the
stochastic variables in the state $\{\rho, v\}$ in the Nelson picture,
and $\Delta(\cdot)$ denote the variances.

{}From eq.(22) it follows that in the Nelson picture the Heisenberg
uncertainty product
$\Delta \hat{q}^{2} \Delta \hat{p}^{2} \;\equiv\; H[\rho,v]$, a
functional
of the density $\rho$ and of the current velocity $v$:
\begin{equation}
H[\rho,v]\;=\; m^{2}\Delta q^{2} \Delta u^{2} \, + \, m^{2}\Delta q^{2}
\Delta v^{2} \, .
\end{equation}

Apart from the constant factor $m^{2}$, the
first term in the right hand side of eq.(23) is the osmotic
uncertainty functional $F[\rho]$ introduced
in section 2, while the second term $m^{2}\Delta q^{2}\Delta v^{2}$
is the current uncertainty functional.

We now have the choice to either extremize only the osmotic term
$F[\rho]$ under variations $\delta \rho$, or
the total uncertainty functional $H[\rho,v]$
under variations $\delta \rho$ and $\delta v$.
In the former case,
one obtains the Gaussian structure (10)-(12), and determines the
current velocity through the continuity equation (17).
The result is
\begin{equation}
v(x,t) \; = \; \langle v \rangle \, + \,
[x -\langle q \rangle ] \: \frac{d\ln\Delta q}{dt} \; .
\end{equation}

 If we take $\Delta q$ to be constant in eq.(24), we have
$v=\langle v \rangle$, and $\Delta v=0$.
The quantum uncertainty then collapses to the minimum
osmotic uncertainty $\hbar^{2}/4$, i.e. to the standard absolute Heisenberg
minimum. The corresponding states have Gaussian density (12) and
phase exponent of the form $\langle \hat{p} \rangle x$: they are the
harmonic-oscillator coherent and squeezed states, the potential
being determined via the HJM equation (21).

With $\Delta q$ time-dependent we have that $\Delta v^{2}
= (d\Delta q/dt)^{2}$. It is straightforward to show that
it is the Schr\"{o}dinger position-momentum correlation:
\begin{equation}
\frac{d}{dt}\Delta q \; = \; \frac{1}{m\Delta q}\,\big[ {\langle \hat{q}
\, \hat{p}
\rangle}_{symm} \; - \; \; \langle \hat{q}\rangle \langle \hat{p} \rangle
\big]
\, ,
\end{equation}

\noindent where ${\langle \hat{q} \, \hat{p} \rangle}_{symm} \; \equiv \;
\langle \hat{q} \, \hat{p} + \hat{p} \, \hat{q} \rangle /2$. We thus have
\begin{equation}
\Delta \hat{q}^{2} \Delta \hat{p}^{2} \; = \; \frac{\hbar^{2}}{4} \; + \;
\big[ {\langle \hat{q} \, \hat{p} \rangle}_{symm} \; - \;
\langle \hat{q}\rangle \langle \hat{p} \rangle \big]^{2} \, .
\end{equation}

{}From the above expressions we see that minimum osmotic uncertainty
with time-dependent $\Delta q$ yields the Schr\"{o}dinger minimum of
the Heisenberg uncertainty product \cite{nietojackiw}.
For the detailed structure
of coherent and squeezed states in stochastic mechanics we refer to
the work by De Martino {\it et al.} \cite{demartino}.

\vspace{0.5cm}

{\bf 4. Kinematical constraint and absolute minimum uncertainty}

\vspace{0.2cm}

So far, it has been shown that minimizing the osmotic uncertainty
functional $F[\rho]$ without constraints reproduces the whole
minimum-uncertainty structure of quantum mechanics provided one exploits,
after minimization has been carried out, the kinematics of diffusions
and the Schr\"{o}dinger dynamics of quantum states.

On the other hand, a fundamental aspect of the calculus of variations regards
the implementation and the meaning of constraints. Minimizing a free
functional and then requiring compatibility of the obtained results
with some constituent equations is in general not equivalent to
minimizing the same functional with the prescribed
equations added as constraints.
True, it is meaningless to seek a constrained minimization
of the osmotic uncertainty $F[\rho]$, since it is a quantity defined
and evaluated at each fixed instant of time. However, wishing to
implement the kinematical and dynamical
equations as constraints, one has simply
to replace $F[\rho]$ with its time-integral over an arbitrarily fixed
interval $[t_{1},t_{2}]$; this new functional is now dependent
on the dynamics: it gives the time-evolution
of the osmotic uncertainty along the process, and it is then sound to
seek its constrained minimization.

We begin by showing that the crucial aspects of uncertainty stem
from the osmotic contribution alone.
To this end, we first
minimize the total uncertainty functional $H[\rho,v]$.
In this case the need for constrained minimization
is imposed {\it ab initio}. Namely, $\rho$ and $v$ are linked by the
continuity and HJM equations, so that it is impossible to consider
independent variations $\delta \rho$ and $\delta v$.

In all generality, one should add both the continuity and the HJM equations
to $H$ and minimize the resulting constrained functional. However, we start
by attacking the simpler problem of minimizing $H$ with the kinematical
constraint alone. In this way dynamics is exploited
after minimization, to select the quantum mechanical processes that
are compatible with the variational results.

Again, when dealing with time-evolution equations, the quantities
to be minimized are not the original functionals, but their time-integrals
over an arbitrary interval. Keeping as
before the normalization constraint,
one has to consider the following functional of $\rho(x,t)$ and $v(x,t)$:
\begin{equation}
{{\cal{H}}}[\rho,v] \equiv \int_{t_{1}}^{t_{2}}dt\,
m^2 {\Delta q}^2 \,( {\Delta u}^2 + {\Delta v}^2 ) +
\int_{t_{1}}^{t_{2}}dt \int_{-\infty}^{+\infty}dx
\bigg\{ \lambda_{0}(t)\rho +
\lambda_{1}(x,t)[\partial_{t}\rho + \partial_{x}(\rho v)]
\bigg\} \: ,
\end{equation}

\noindent with the arbitrary functions $\lambda_0(t)$ and $\lambda_1(x,t)$
as Lagrange multipliers.

By requiring stationarity of ${\cal H}$ with respect to independent
variations of $\rho$ and $v$ we are left with the following equations:
\begin{equation}
2\,m^2{\Delta q}^2 (v-\langle v\rangle) - \partial_{x}\lambda_1 \,= \,0\;,
\end{equation}
\begin{equation}
m^{2}({\Delta u}^2+{\Delta v}^2)(x^2-2\langle q \rangle x)
+ m^2{\Delta q}^2
\bigg( v^2-2 \langle v \rangle v -
\frac{\hbar^2}{m^2} \frac{ \partial_{x}^2 \sqrt{\rho} }
{ \sqrt{\rho} } \bigg)
+ \lambda_0 (t)- \partial_{t}\lambda_1-v\partial_{x}\lambda_1
\,=\,0 \;\,.
\end{equation}

The unknown multipliers $\lambda_0$ and $\lambda_1$ can be
eliminated from these relations by inserting eq.(28) and its time
derivative into the space derivative of eq.(29).
Differentiation with respect to time obviously
yields two distinct results according to whether the spreading $\Delta q$ is
taken to be constant or time-dependent.

In the first case one has, recalling the definition of $u$, eq.(6),
\begin{equation}
\partial_{t}(v-\langle v \rangle)+v\partial_{x}v - \frac{ {\Delta u}^2
+{\Delta v}^2 }
{{\Delta q}^2}(x-\langle q \rangle) +
\frac{1}{2} \partial_{x}\bigg( u^2+\frac{\hbar}{m}\partial_{x} u\bigg)
\,=\,0  \:.
\end{equation}

For the Nelson diffusion of quantum mechanics the above equation must hold
together with the HJM dynamics, eq.(21). One can show, after some
manipulations, that a sufficient condition for the two equations to be
consistent is $v=\langle v \rangle$.
This condition reduces eq.(30) to the form
\begin{equation}
\partial_{x}\left\{
{\Delta u}^2 (x^2-2\langle q \rangle x)-
\bigg(u^2+\frac{\hbar}{m}\partial_{x} u\bigg) {\Delta q}^2 + \lambda(t)
\right\}\,=\,0\,.
\end{equation}

This is just the space derivative of eq.(6) of section 2, with $\Delta q$
now fixed to be constant, and we again
recover the Heisenberg minimum uncertainty states.

In the case of time-dependent $\Delta q$, one has instead, from eqs.(28)-(29),
\begin{eqnarray}
\partial_{t}(v-\langle v \rangle)+2(v-\langle v \rangle)
\,\partial_{t}\ln \Delta q
&\hspace{-1mm} + \hspace{-1mm}& v\partial_{x}v
- \frac{ {\Delta u}^2 + {\Delta v}^2 }{ {\Delta u}^2 }
( x-\langle q \rangle )\nonumber \\
\mbox{}&\hspace{-1mm} +\hspace{-1mm} & \frac{1}{2}\partial_{x}
\bigg( u^2+\frac{\hbar}{m}\partial_{x} u\bigg) \,=\,0  \,.
\end{eqnarray}

The problem is now slightly more complicated, but one can convince himself,
proceeding as before, that HJM dynamics is compatible with the choice of the
linear $v$, eq.(24). By inserting this form of $v$ in eq.(32),
the linear $u$ of minimum uncertainty is recovered if the
time-evolution of $\Delta q$ is ruled by
\begin{equation}
\frac{{d}^2}{{dt}^2}\,\Delta q+\frac{1}{\Delta q} \,
{\bigg( \frac{d}{dt}\,\Delta q \bigg) }^2\,=\,0\; .
\end{equation}

One can realize by direct computation that the above expresses
the condition for the current uncertainty functional
${\Delta q}^2{\Delta v}^2$, namely the Schr\"{o}dinger part of the uncertainty
product, to be a constant of the motion.
Eq.(33) can be explicitely solved to give
\begin{equation}
\Delta q(t)\,=\,{\Delta q}_0\, \sqrt{{1+\beta t}}\;,
\end{equation}
\noindent where ${\Delta q}_0 = \Delta q(t=0)$,
$\beta = 2\,{( d\Delta q/dt )}_{t=0}/\Delta q(0)$; consequently,
\begin{equation}
m^2{\Delta q}^2{\Delta u}^2+m^{2}{\Delta q}^2{\Delta v}^2\,=\,
\frac{\hbar^2}{4}+ m^2 {\left[ {\Delta q}^2 {( d\Delta q/dt )}^2
\right] }_{t=0}\;.
\end{equation}

The corresponding normalized solutions are states of time independent
Schr\"{o}dinger minimum. They are characterized by a wave packet
width that evolves according to the diffusive dispersion law Eq.(34).

We see that constrained minimization of the total uncertainty functional
selects all the states of Heisenberg minimum uncertainty, but only the
special class of constant Schr\"{o}dinger minima. This is in contrast
with the results obtained extremizing the free osmotic functional which
allowed to derive all the states of minimum quantum uncertainty.
For this reason, we consider again the osmotic uncertainty and seek for its
constrained minimization. Proceeding as before, we introduce the functional
\begin{eqnarray}
{\cal C}[\rho,v] & \equiv & \int_{t_{1}}^{t_{2}}dt\,
m^2 {\Delta q}^2  {\Delta u}^2 \nonumber \\
                 &    +   &
\int_{t_{1}}^{t_{2}}dt \int_{-\infty}^{+\infty}dx
\bigg\{ \lambda_{0}(t)\rho +
\lambda_{1}(x,t)\,[\partial_{t}\rho + \partial_{x}(\rho v)]
\bigg\} \, .
\end{eqnarray}

Performing the independent variations of $\rho$ and $v$ and requiring
stationarity of ${\cal C}$ leads to a couple of equations that replace
eqs.(28)-(29)
\begin{equation}
\partial_{x}\lambda_1\,=\,0\;,
\end{equation}
\begin{equation}
m^2{\Delta u}^2(x^2-2\langle q \rangle x)
- {\hbar^2}{\Delta q}^2
  \frac{ \partial_{x}^2 \sqrt{\rho} } { \sqrt{\rho} }
 + \lambda_0 (t)- \partial_{t}\lambda_1-v\partial_{x}\lambda_1
\,=\,0 \;\,.
\end{equation}

\noindent Eliminating the unknown multiplier $\lambda_1$ from the equations
one is left with
\begin{equation}
\partial_{x}\left\{
{\Delta u}^2 (x^2-2\langle q \rangle x)-
\bigg(u^2+\frac{\hbar}{m}\partial_{x} u\bigg) {\Delta q}^2 + \lambda(t)
\right\}\,=\,0\,.
\end{equation}

We are again in the general case of section 2, eq.(6). We remark that now,
contrary to the situation faced in the derivation of eqs.(30)-(31), no
hypotheses need to be made on the time dependence of $\Delta q$.
In conclusion, we recover all the states of Heisenberg and
Schr\"{o}dinger minimum uncertainty. Thus,
the complete structure of minimum quantum uncertainty is
derived by considering the osmotic contribution
as the correct functional to be extremized.

\vspace{0.5cm}

{\bf 5. Discussion and outlook}

\vspace{0.2cm}

The analysis carried out in the previous sections implies two
main considerations. The first one is the emergence of the crucial
role played by the osmotic uncertainty product in determining
the fundamental features of the quantum mechanical uncertainty.

The other main implication is that the variational approach lends
itself naturally to a further generalization.
Namely, it would be very interesting to exploit the dynamics directly
as a constraint in the variational procedure
to minimize the osmotic uncertainty. The physical motivation for
such a construction is that it might allow to discuss the behaviour
and the structure of quantum uncertainty for general non-harmonic
systems. For instance, it would be interesting to verify whether
the Heisenberg uncertainty product exhibits local minima depending
on the choice of the potential.

To implement the suggested generalization one should consider
the functional
\begin{eqnarray}
\overline{\cal C}[\rho,v] & \equiv & \int_{t_{1}}^{t_{2}}dt\,
m^2 {\Delta q}^2  {\Delta u}^2   \nonumber \\
                    &   +    &
\int_{t_{1}}^{t_{2}}dt \int_{-\infty}^{+\infty}dx
\bigg\{ \lambda_{0}(t)\rho +
\lambda_{1}(x,t)\,[\partial_{t}\rho + \partial_{x}(\rho v)] \nonumber \\
                    &   +    &
\lambda_{2}(x,t) \bigg[ \partial_{t}v + v\partial_{x}v
- \frac{\hbar^2}{2m^2} \partial_{x}\frac{ \partial_{x}^2 \sqrt{\rho} }
{ \sqrt{\rho} } + \frac{ \partial_{x}V }{m} \bigg]
\bigg\} \, .
\end{eqnarray}

The above defines the most general osmotic uncertainty functional in
quantum mechanics, including the normalization, the kinematical and
the dynamical constraints.
Extremizing it as usual against smooth independent variations of
$\rho$ and $v$ one obtains, as in the kinematically
constrained scheme, two coupled equations. However, in this case
the difficulty of identifying and/or eliminating the unknown Lagrange
multipliers has prevented so far an explicit exact solution of the
problem. Work is in progress, to obtain both numerical and analytical
approximated solutions, and we will report on it elsewhere \cite{illv}.

\vspace{0.5cm}

{\bf Acknowledgement}

\vspace{0.2cm}

We wish to thank Prof. Laura M. Morato for suggesting us the
study of this very stimulating problem, and for being a constant
source of help and advice.

\end{document}